\documentclass[12pt]{article}
\usepackage{pdproc,epsfig}
\begin{document}

\renewcommand{\thefootnote}{\alph{footnote}}
  
\title{Supernova Neutrinos\footnote{Invited talk at the
X International Workshop on
Neutrino Telescopes, Venice, Italy, March 11-14, 2003,}}

\author{ Petr Vogel}

\address{ 
 Institute of Particle and Nuclear Physics, Charles University, Prague, Czech Republic, \\
on leave from Physics Department, Caltech, Pasadena, CA, USA\\
 {\rm E-mail: pxv@caltech.edu}}

\abstract{I describe how the signals corresponding to the supernova $\nu_e$ and
$\bar{\nu}_e$ (charged current reactions)
as well as all active neutrinos (neutral current reactions) 
can be separately observed in various existing detectors. 
These observations would make it possible 
to determine the flux and average energy (or temperature)
for each of these three neutrino signal components. 
I argue that all these
quantities are needed in order to understand the interplay between the
so far poorly
known supernova neutrino emission process and the neutrino oscillations.}
   
\normalsize\baselineskip=15pt

\section{Introduction}
Supernovae are fascinating objects. The core collapse represents
one of the most energetic events known and the explosion that propels
the envelope into the interstellar space and ultimately causes the visible
effects noted already in historic times is poorly understood.
But most of the gravitational energy of the collapse is not 
emitted as the
kinetic energy of the outward explosion, and even less as the visible
light. Instead, it is carried away
by neutrinos. In fact, it is curious to note that, when averaged over
a sufficiently long time ($>$ 100 years), the energy of the emitted
neutrinos is comparable to the electromagnetic energy emitted
by the whole galaxy over that time.

It is also likely that the hot nucleon gas pushed away 
by the neutrinos from the newly
born proto-neutron star is the site of the $r$-process where roughly
half of the heavy elements are produced. Understanding the neutrino
emission process following the core collapse is thus a necessary
step for the understanding of the $r$-process nucleosynthesis.

At the same time observation of supernova neutrinos represents probably
the longest baseline neutrino oscillation experiment possible. Neutrinos
emitted from the proto-neutron star might undergo matter oscillations
while passing through the atmosphere and envelope of the star, and vacuum
oscillations on the way to Earth. Finally, before reaching the detector
they might be influenced by the passage through Earth. Thus, the detected
neutrinos are potentially a complicated convolution of the primary
fluxes reflecting the initial production inside the star, and
the modifications due to oscillations on their way to the detectors.
   
It is therefore obvious that observation of neutrinos emitted by a future
supernova in our galaxy is going to be a goldmine of useful information.
However, supernovae in our galaxy are very rare events, and supernova
neutrino fluxes are observable for less than a minute per century
(duty cycle of about $10^{-8}$). It is therefore important to anticipate
what might happen, and make every effort to maximize the information
gleaned from such a rare occasion.
Each physicist will have, at best, only one chance to observe neutrinos
from the galactic supernova during his or her scientific career.
(Unfortunately, the present or planned neutrino detectors are unable
to observe neutrinos with non-negligible statistics associated with
supernovae even in the nearest galaxy, Andromeda, about 700 kpc
away.)

As ``guiding principles'' for the modelling of the neutrino emission
following the core collapse we will use four rules:
\begin{enumerate}
\item Essentially all gravitational binding energy, 
$E_B \sim G_N M_{\odot} / R$
 ($G_N$ is the gravitational constant, $R \sim$ 10 km 
is the neutron star radius), is emitted in neutrinos. For application
we use $E_B = 3 \times 10^{53}$ ergs, and the distance of 10 kpc
(about half of the stars in our galaxy are within that distance).
\item The characteristic neutrino emission time, related to the neutrino
diffusion time, is $\sim$ 10 seconds.
\item `Equal luminosity rule' states that the total emitted energy is
equally shared by all six neutrino flavors. Thus, the typical luminosity
of each neutrino flavor is $\sim 5 \times 10^{51}$ erg/s.
\item `Temperature hierachy rule' states that, as a consequence of different
cross sections for $\nu_e, \bar{\nu}_e$ and $\nu_x \equiv \nu_{\mu}, 
\nu_{\tau}, \bar{\nu}_{\mu}, \bar{\nu}_{\tau}$ the average energies
are \underline{not} equal and the hierarchy 
$\langle E_{\nu_e} \rangle <  \langle E_{\bar{\nu}_e} \rangle <
\langle E_{\nu_x} \rangle$ is expected. For applications we use
the values $\langle E_{\nu_e} \rangle =$ 11 MeV, 
$\langle E_{\bar{\nu}_e} \rangle =$ 16 MeV, 
$\langle E_{\nu_x} \rangle =$ 25 MeV.
\end{enumerate}

Supernova neutrinos, presumably only $\bar{\nu}_e$, were detected
so far only once, 26 years ago, when the neutrino signal of SN1987A was 
observed\cite{kam87,imb87}. That signal, taking into account its limited
statistics, confirmed some of the rules above, but naturally 
told us nothing about the fluxes and average energies of the unobserved
neutrino flavors. 

The above rules, in particular the items 3) and 4) are based on 
simulations\cite{Woosley,Totani} of the neutrino emission process. 
However, recent studies\cite{Raffelt} point out that these rules
are highly model dependent, and that relatively large deviations
from both of them might be expected.     
  
Supernova neutrinos can be detected on Earth using three reaction types.
The charged current reactions with the production of $e^+$,
changing a free or bound proton into a free or bound neutron,
are sensitive only to the $\bar{\nu}_e$ component. Similarly,
the charged current reactions with the production of $e^-$,
changing a  bound neutron into a free or bound proton,
are sensitive only to the $\nu_e$ component. Finally, the neutral current
reactions are sensitive to \underline{all} active neutrino flavors.
(Neutrino-electron scattering is caused by a combination of the charged
and neutral current reaction amplitudes. Observationally, one cannot
distinguish the initial neutrino flavor in that case.)

In the ideal situation one would be able to determine separately
and with sufficient statistical accuracy the rate of each of these
reactions. Moreover, spectroscopic information (at least the extraction
of the average energy or equivalently temperature) should be possible.
And, still in the ideal situation, all this information should be
extracted as a function of time. 

In a less ideal and a bit more realistic case, an average over time of the
six quantities, the fluxes and average energies of the three components,
should be extracted from observations.

There are, at present, several operational detectors of a sufficient
size to detect the galactic supernova neutrino signal (SuperKamiokande,
SNO, KamLAND, LVD, AMANDA, and MiniBoone;
see other talks at this conference for the description
of these detectors). Several other detectors
are being built or are planned (e.g. Borexino, OMNIS, LAND,
Nestor, Antares, IceCube). All, or most, of
these detectors were or are going to be
built for a different purpose; observation of supernova
neutrinos will be in a parasitic mode. Most of them have a `Supernova trigger',
i.e. a piece of software and/or hardware that is designed to warn
the observers that a signal resembling a supernova was detected, and 
possibly switching thresholds etc. to maximize that signal.
There is an international collaboration of supernova neutrino
detectors (SNEWS, SuperNova Early Warning System, see e.g. \cite{Kate})
to provide, with high confidence, an early alert from the coincidence
of neutrino signals in several detectors.  

In the following I concentrate on three existing detectors:
SuperKamiokande is a 32 kton water \v{C}erenkov detector with
a detection threshold of about 5 MeV, SNO is a 1 kt heavy water 
\v{C}erenkov detector, again with a threshold of about 5 MeV,
and KamLAND is again a 1 kt but liquid scintillator detector with a much
lower threshold, of only few hundred keV.

In each of these detectors supernova neutrinos will be observed
in several ways. Below I will explain how the combination of these
signals, with the proper arrangements of the triggers,
should be able to determine, with a reasonable accuracy, 
the six parameters described above, i.e. the flux and temperature 
of the three expected components of the core collapse 
supernova neutrino signal.

\section{Charged current reactions}
Detecting $\bar{\nu}_e$ neutrinos is relatively easy. All considered
detectors contain free protons, and the inverse neutron beta decay
\begin{equation}
\bar{\nu}_e + p \rightarrow e^+ + n
\end{equation}
has large and well understood cross section\cite{cross} (accuracy
of $\sim$ 0.2\%). By measuring the positron energy one can deduce 
the incoming $\bar{\nu}_e$ energy, since
\begin{equation}
E_{\bar{\nu}_e} \simeq E_{e^+} + M_n - M_p ~.
\end{equation}
In SuperKamiokande one expects about 8000 events of this type from
the `standard' supernova. Thus, it would be possible to determine the
flux and average energy of this component with a good accuracy in several
time bins. 

The other considered detectors, SNO and KamLAND, will detect several
hundred events of this type each (in SNO due to the light water part
of the detector). In addition, in SNO, the reaction 
\begin{equation}
\bar{\nu}_e + d \rightarrow e^+ + n + n
\end{equation}
will have about 80 events\cite{BVII}. 
Their identification depends on the
way and efficiency of the neutron detection at SNO
employed at the time of the supernova detection.

It is more difficult to detect $\nu_e$. Since there are
no free neutrons, the detection reaction must be based
on neutrons bound in a nucleus. The corresponding cross
sections are smaller than for the inverse neutron beta decay, eq.(1).
Moreover, with few notable exceptions (deuteron, reaction on
$^{12}$C populating the ground state of $^{12}$N), these cross
sections have not been measured and thus their value is based
on calculations involving nuclear models. That introduces some
uncertainty into the deduced quantities.

In SNO the deuteron disintegration 
\begin{equation}
\nu_e + d \rightarrow e^- + p + p
\end{equation}
will result also in about 80 events\cite{BVII}. They have to be
distinguished, by their lack of neutrons, from the corresponding
reaction with $\bar{\nu}_e$, eq.(3).  

In KamLAND there will be a handful of clean events 
of the type $\nu_e + ^{12}$C $\rightarrow e^- + ^{12}$N$_{gs}$.
Such events are easy to recognize since N$_{gs}$ decays 
by the $\beta^+$ emission with
half-life of 11 ms back to carbon. If, as a result of oscillations,
the $\nu_e$ average energy is considerably larger than our guess,
there will be correspondingly larger number of such events. However,
it might be difficult to separate them from the mirror reaction
$\bar{\nu}_e + ^{12}$C $\rightarrow e^- + ^{12}$B$_{gs}$
with a similar signature and yield.

In SuperKamiokande the reaction $\nu_e + ^{16}$O $\rightarrow
e^- + ^{16}$F$^*$ results in only about 20 events when the 
`standard' $\nu_e$ average energy is assumed. However, again if through 
oscillations the effective $\nu_e$ temperature would increase
to $T_{\nu_e}$ = 8 MeV, the yield would increase dramatically,
to $\sim$860 events\cite{haxton}. The angular distribution of these
electrons is rather different that the 
distribution of positrons from the inverse
neutron beta decay. This feature could be used in order to
separate these two channels.

Finally, I should mention plans to develop a supernova lead
based neutrino detector. In it, one would 
count either the number of neutrons emitted in the charged
or neutral current reactions leading to the continuum in
the final nuclei; or observe the electrons from the charged
current reactions as well. Several recent publications have
been devoted to the theoretical prediction of the corresponding
cross sections \cite{Pb1,Pb2,Pb3,Pb4}. The spread between these
calculated cross sections illustrates the difficulties encountered when
dealing with the neutrino induced reactions on complex nuclei.
 
Even though the relation between the incoming $\nu_e$ energy 
and the outgoing electron energy is not as simple as in eq.(2)
since one has to take into account the spread of energies 
of the final nuclear system, it is reasonable to assume 
that one or several of the above ways of detecting the $\nu_e$
signal would make it possible 
to check the `hierarchy rule' stated above,
or to conclude that, presumably due to neutrino oscillations,
the $\nu_e$ component on Earth has considerably higher energy
than usually expected.

\section{Neutral current reactions}
Neutral current reactions measure the flux of all active neutrinos.
Typically, neutral current cross section are increasing
functions of energy. Thus, if the `temperature hierarchy' and
`equal luminosity' rules are valid, one expects that the
yield of the neutral current reactions will be dominated
by the $\nu_x$ neutrinos. (These neutrinos are in that case 
responsible for 4/6 of the total luminosity, and have higher
energy, and hence bigger yield per particle.) Thus, in some sense,
the yield of the neutral current reaction is a measure of the
contribution of the $\nu_x$ neutrinos to the supernova luminosity.

There are several observable neutral current reactions in the considered
detectors:
\begin{enumerate}
\item Neutrino-electron scattering has observable rate in all detectors.
However, as mentioned earlier, it is difficult to separate the 
neutral current part, since the charged current is usually dominating.
\item In water one can observe the $\gamma$ rays following the
inelastic scattering of neutrinos on oxygen\cite{LKV}. 
\item In heavy water (SNO) the neutral current deuteron disintegration,
$\nu + d \rightarrow \nu + p + n$,
can be detected by counting the number of singly produced neutrons.
SNO collaboration has demonstrated that they mastered this technique
\cite{SNOn}. 
\item Inelastic neutrino scattering on a heavy nucleus often results
in excitation of the continuum, followed by the emission of one or
more neutrons, e.g. $\nu + ^{208}$Pb $\rightarrow \nu' + ^{207}$Pb $+ n$.
Again, the detection will be based on counting the number of produced
neutrons.
\item In a liquid scintillator detecor, which contains carbon, the
excitation of the $T = 1, I^{\pi} = 1^+$ state in $^{12}$C
results in the emission of the clearly recognizable $\gamma$ line 
at 15.11 MeV.
\item  In a detector with sufficiently low thresholds (KamLAND, Borexino)
the elastic scattering on protons $\nu + p \rightarrow \nu' + p'$
can be observed by detecting the recoiling protons.
\end{enumerate}

It is important to note that the neutral current reactions in items 2-5
have \underline{no spectral information}. One measures simply the number
of events
\begin{equation}
N_{nc} \sim \frac{1}{D^2 \langle E_{\nu} \rangle} 
\int f(E_{\nu}) \sigma(E_{\nu}) dE_{\nu} ~,
\end{equation}
where $D$ is the supernova distance, $f(E_{\nu})$ is the neutrino spectrum,
and $\sigma(E_{\nu})$ is the cross section. Thus, there is
a parameter degeneracy. Higher average energy and correspondingly lower
flux can produce the came number of events. It is difficult 
to overcome this problem.

\begin{figure}[h]
\vspace*{13pt}
         \mbox{\epsfig{figure=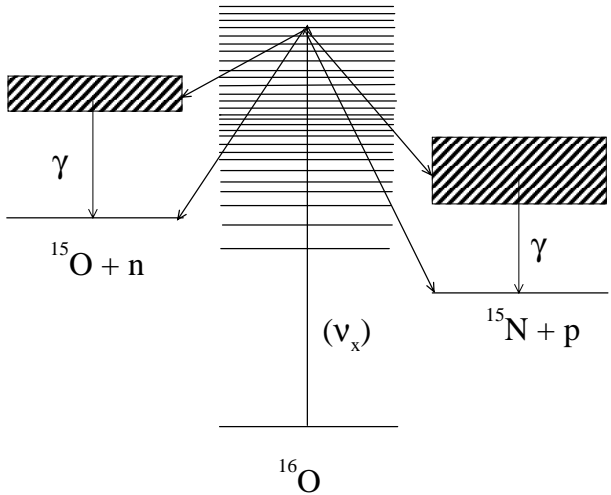,width=8.0cm}}
\caption{Schematic illustration of the detection scheme for the
neutral current detection in water \v{C}erenkov detectors.}
\label{fig:o16}
\end{figure}

The detection of the neutral current inelastic scattering on $^{16}$O
deserves an explanation. The principle is illustrated in Fig. \ref{fig:o16}.
When neutrinos of a sufficient energy inelastically scatter on $^{16}$O,
the resulting excited state decays by nucleon emission, leading to
the ground or excited states of $^{15}$N (proton emission) or
$^{15}$O (neutron emission). These mirror nuclei have only bound excited
states with energies between 5 and 10 MeV. Thus, every time the
excited state in one of these two nuclei is reached (branching of
$\sim$30\%) a photon of energy of at least 5 MeV, detectable in SK,
is emitted. The yield of this reaction is particularly sensitive to
the high energy tail of the neutrino spectrum, i.e. to the deviation
(pinching) from the Fermi-Dirac shape\cite{LKV}.

The elastic scattering of neutrinos on protons (item 6) above)
is the only neutral current reaction that, at least in principle,
can furnish spectroscopic information among the neutral current
reactions considered above\cite{will}. 
(A complementary information could be,
perhaps, extracted from the $\nu - e$ scattering data, after 
a statistical subtraction of the $\nu_e - e$ and  $\bar{\nu}_e - e$
events.)
 
\begin{figure}[h]
\vspace*{13pt}
         \mbox{\epsfig{figure=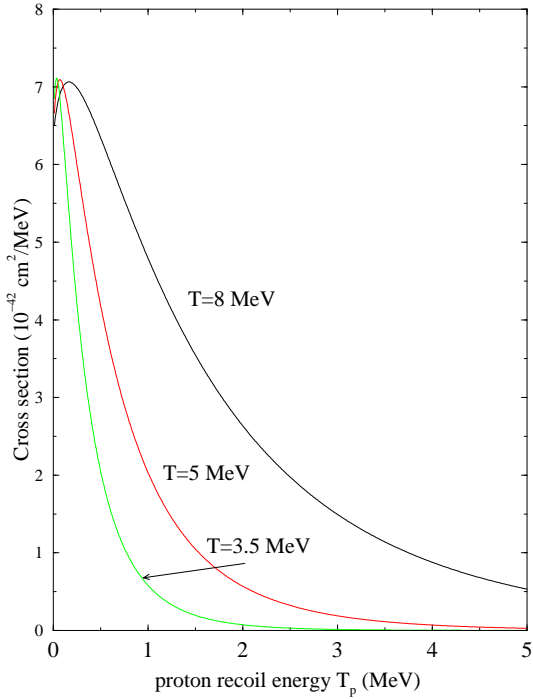,width=8.0cm}}
\caption{Cross section of the elastic neutrino scattering
on protons for the indicated incoming neutrino temperatures.
True proton recoil energy, without quenching, is used}
\label{fig:fig3}
\end{figure}

The cross section for $\nu - p$ elastic scattering is well understood.
For $E_{\nu} \ll M_p$ it is given by ($T_p$ is the recoil proton
kinetic energy) 
\begin{equation}
\frac{d \sigma}{d T_p} \simeq \frac{G_F^2 M_p}{\pi} \left[
(c_A^2 + c_V^2) + (c_A^2 - c_V^2)\frac{T_p M_p}{E_{\nu}^2} \right]
~,~~c_V = 1/2 - 2\sin^2 \theta_W, ~~c_A = 1.27/2 ~.
\end{equation}
However, the proton recoil energy is quite small, restricted to
$T_p < 2 E_{\nu}^2/M_p$. The recoiling proton is obviously
nonrelativistic, and can be detected only by its ionization.
Moreover, in a liquid scintillator the light yield of heavy 
ionizing particles is reduced (quenched) when compared to the
light yield of the electrons or photons. The cross section, 
integrated over the Fermi-Dirac energy distribution, is shown
in Fig. \ref{fig:fig3}. Note the steep dependence on the
neutrino temperature. 

Obviously, only detectors with sufficiently low threshold ($\ll$ 1 MeV)
of the the `effective' (i.e. quenched) proton recoil energy
and correspondingly low background at those energies
could be used. According to Ref.\cite{will} one expects about
300 events per kt above 200 keV. Thus, the background at these
energies should be at most in the few Hz range. Assuming that
these extremely stringent conditions could be met (both KamLAND
and Borexino plan to meet them), it would be possible
not only to detect the recoil protons but to extract useful
spectroscopic information out of that signal.

\begin{figure}
\vspace*{13pt}
         \mbox{\epsfig{figure=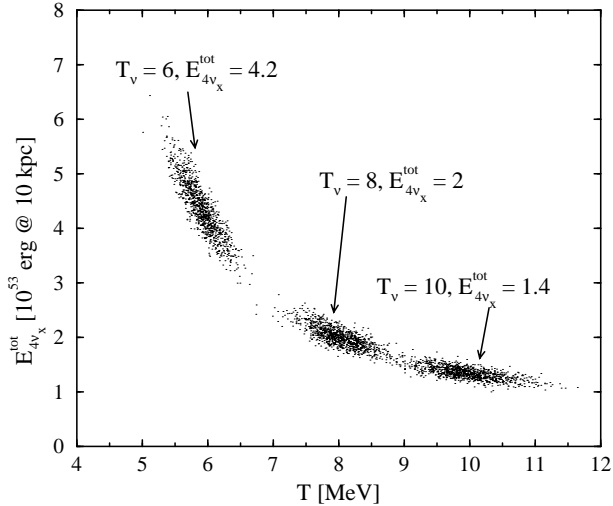,width=8.0cm}}
\caption{Monte Carlo simulation of the combined fit to 
$T_{\nu_x}$ and the total
energy carried by such neutrinos, $E_{4\nu_x}^{tot}$.}
\label{fig:will}
\end{figure}

This is illustrated in Fig. \ref{fig:will} where it is demonstrated that
the parameter degeneracy between the luminosity of the $\nu_x$ neutrinos 
and their temperature can be resolved. The values of the total energy
of the $\nu_x$ neutrinos and their temperature were chosen for this
illustration in such a way that the total number of events above threshold
is the same. 

\section{Neutronization pulse}

In the early stages of the core collapse electrons are captured
on the core nuclei, and the resulting $\nu_e$  escape. Eventually
the density of the core increases so much that the neutrinos
are trapped. However, outside the outgoing
shock the $\nu_e$ created by the electron captures still escape, forming
a very narrow ($\sim$0.01 s) and intense pulse. This pulse represents
$2 - 5 \times 10^{51}$ ergs of energy, i.e. perhaps $5 - 10$\% of the
total energy of the $\nu_e$ neutrinos (see, e.g. Ref.\cite{pulse}).  

Scaling the yield estimates in Ref.\cite{BVI} I estimate that
such a pulse could result in $\sim$ 10 $\nu_e - e$ scattering events in SK,
recognizable by their clustering in time, and by their characteristic
narrow angular distribution. (Obviously, that yield estimate has
a substantial error margin.) 

Clearly, observation of the neutronization pulse would be valuable
for understanding the mechanism of the core collapse and bounce.

\section{Relic supernova neutrinos}
Neutrinos emitted by the core collapse supernovae move freely and accumulate
over time. Eventually, they form a diffuse (in time and direction) 
neutrino flux, when averaged over $\sim10^8$ galaxies ($10^8$ is the ratio
of the time elapsed between supernovae 
per galaxy $\sim$30 years, and the neutrino
pulse duration $\sim$ 10 seconds.) That diffuse flux is cut-off naturally
at the redshift $z \sim 1$, since the neutrinos emitted at higher
redshifts become unobservable due to their lower energy.
Crude estimate of this diffuse flux suggests that its $\bar{\nu}_e$
component is $\sim10 \bar{\nu}_e$/(cm$^2$ s) resulting in about
0.5 events per kt and year. 

More detailed evaluations basically confirm that estimate\cite{Gary,TS}.
Observation of such diffuse flux would be a measure of the average
SN rate over a substantial fraction of the Universe. 

Clearly, the $\bar{\nu}_e$ component of the diffuse flux has the
best chance to be seen. There is a `window of opportunity' for its
observation, above the energies of the reactor and solar neutrinos,
and below the energy of the atmospheric neutrino signal.
Both Kamiokande\cite{Kam_d} and SuperKamiokande\cite{SK_d} reported
limits on the diffuse flux. The much better SK limit, 
which is background limited, is 
approaching (but still well above) the theoretical estimate
of this flux. It appears that one needs a much larger detector,
with correspondingly smaller background per unit mass, to be able
to positively identify this important quantity.
  
\section{Acknowledgements}
  This work  was  supported in part
by the Center for Particle Physics, project No. LN00A 006
of the Ministry of Education of the Czech Republic
and by the US Department of Energy Grant DE-FG03-88ER40397.


\begin{thebibliography}{99}
\bibitem{kam87} K. Hirata {\it et al.}, {\it Phys. Rev. Lett.} {\bf 58}
(1987),1490. 
\bibitem{imb87} R. Bionta {\it et al.}, {\it Phys. Rev. Lett.} {\bf 58}
(1987),1494.
\bibitem{Woosley} S. E. Woosley  {\it et al.},{\it Astrophys. J} {\bf 433} 
(1994) 229. 
\bibitem{Totani} T. Totani, K. Sato, H. E. Dalhed, and J. R. Wilson,
{\it Astrophys. J} {\bf 496} (1998) 216.
\bibitem{Raffelt} M. Th. Keil, G. Raffelt and H-T. Janka, astro-ph/0208035.
\bibitem{Kate} K. Scholberg, {\it Nucl. Phys. Proc. Suppl.} {\bf 91}
(2000) 331.
\bibitem{cross} P. Vogel and J. F. Beacom,   {\it Phys. Rev.D} {\bf 60}
(1999) 053003.
\bibitem{BVII} J. F. Beacom and P. Vogel, {it Phys. Rev.D} {\bf 58} 
(1998) 093012.
\bibitem{haxton} W. C. Haxton,  {\it Phys. Rev.D} {\bf 36} (1987) 2283.
\bibitem{Pb1} G. Fuller, W.C. Haxton and G. C. McLaughlin,
{\it Phys. Rev. D}{\bf 59}  (1999) 085005.
\bibitem{Pb2} E. Kolbe and K. Langanke,
{\it Phys. Rev. C}{\bf 63} (2001) 025802.
\bibitem{Pb3} C. Volpe, N. Auerbach, G. Colo, and N. Van Giai,
{\it Phys. Rev. C}{\bf 65} (2002) 033007.
\bibitem{Pb4} J. Engel, G. C. McLaughlin, and C. Volpe, hep-ph/0209267. 
\bibitem{LKV}  K. Langanke, P. Vogel, and E. Kolbe,  {\it Phys. Rev. Lett.} 
{\bf 76} (1996) 2629.
\bibitem{SNOn} Q. R. Ahmad {\it et al.}, {\it Phys. Rev. Lett.} 
{\bf 89} (2002) 011301.
\bibitem{will} J. F. Beacom, W. M. Farr, and P. Vogel,  {\it Phys. Rev.D}
{\bf 66} (2002) 033001.
\bibitem{pulse} R. Buras, M. Rampp, H.-Th. Janka and K. Kifonidis,
astro-ph/0303171.
\bibitem{BVI} J. F. Beacom and P. Vogel, {\it Phys. Rev.D} {\bf 58} 
(1998) 053010.
\bibitem{Gary} M. Kaplinghat, G. Steigman and T. P. Walker, {it Phys. Rev.D} 
{\bf 62} (2000) 043001.
\bibitem{TS} T. Totani and K. Sato, {\it Astrop. Phys.} {\bf 3} (1995) 367.
\bibitem{Kam_d} W. Zhang {\it et al.}, {\it Phys. Rev. Lett.} 
{\bf 61} (1988) 385.
\bibitem{SK_d} M. S. Malek {\it et al.}, {\it Phys. Rev. Lett.} 
{\bf 90} (2003) 061001.
\end{thebibliography}
\end{document}